\documentclass[twocolumn]{aastex62}

\received{XX, 2019}
\revised{XX, 2019}
\accepted{\today}
\submitjournal{ApJLett}

\shorttitle{A new Einstein Cross gravitational lens}
\shortauthors{Bettoni et al.}

\begin{document}

\title{A new Einstein Cross gravitational lens of a Lyman-break galaxy}

\correspondingauthor{Daniela Bettoni}
\email{daniela.bettoni@inaf.it}

\author[ 0000-0002-4158-6496]{Daniela Bettoni}
\author{Renato Falomo}
\affil{INAF - Osservatorio Astronomico di Padova, 
Vicolo Osservatorio 5 35122 Padova, Italy}

\author{Riccardo Scarpa}
\affiliation{GRANTECAN, Cuesta de San Jos\'e s/n, E-38712 , Bre\~na Baja, La Palma, Spain,
 Instituto de Astrof'sica de Canarias, V\'ia L\'actea s/n, E38200, La Laguna, Tenerife, Spain}

\author{Mattia Negrello}
\affiliation{School of Physics and Astronomy, Cardiff University, The Parade, Cardiff CF24 3AA, UK}

\author{Alessando Omizzolo}
\affiliation{Vatican Observatory, Vatican City State, Vatican City}
\affil{INAF - Osservatorio Astronomico di Padova, 
Vicolo Osservatorio 5 35122 Padova, Italy}

\author{Romano L. M. Corradi}
\author{Daniel Reverte}
\affiliation{GRANTECAN, Cuesta de San Jos\'e s/n, E-38712 , Bre\~na Baja, La Palma, Spain,
 Instituto de Astrof'sica de Canarias, V\'ia L\'actea s/n, E38200, La Laguna, Tenerife, Spain}

\author{Benedetta Vulcani}
\affil{INAF - Osservatorio Astronomico di Padova, 
Vicolo Osservatorio 5 35122 Padova, Italy}




\begin{abstract}

We report the study of an "Einstein Cross" 
configuration first identified in a set of HST images by \citet{Cerny}.
Deep spectroscopic observations obtained at 
the Spanish 10.4m GTC telescope, allowed us to demonstrate the lens nature of the system, that consists
of a Lyman-break galaxy, not a QSO as is usually the case, at z = 3.03 lensed by a galaxy at z=0.556. 
Combining the new spectroscopy with the archival HST data, it turns out that
the lens is an elliptical galaxy with M$_V =-21.0$, effective radius 2.8 kpc and stellar velocity dispersion $\sigma$=208$\pm$39 km/sec. The source is a Lyman break galaxy with Ly$\alpha$ luminosity $\sim L^*$ at that redshift.
 From the modeling of the system, performed by assuming a singular isothermal ellipsoid (SIE) with external shear, we estimate that the flux source is magnified about 4.5 times, and the velocity dispersion of the lens is $\sigma_{\rm SIE}=197.9_{-1.3}^{+2.6}$\,km\,s$^{-1}$, in good agreement with the value derived spectroscopically. 
This is the second case known of an Einstein cross  of a Lyman-break galaxy.

\end{abstract}



\section{Introduction} \label{sec:intro}
Gravitational lenses represent one of the most powerful tool to probe the properties of distant galaxies and the cosmological parameters. Strong gravitational lensing produces multiple images of distant sources that have their
line of sight very close to foreground massive objects \citep[see][for a review]{TT10} .  In addition the special case of image splitting of distant quasars can provide the direct measurement
of the Hubble constant from correlated flux variability
\citep{TT2,Suyu14}.

Accurate lens modelling can precisely probe the density profile of galaxies at cosmological distances, specifically the mass enclosed
within the Einstein radius and the mean local
density slope within it. Combining lensing and dynamics allows the
central dark matter profile to be robustly inferred \citep[e.g.][]{TT}. All these facts
make this kind of configuration a fantastic laboratory for the study
of the Universe.

Of particular interest is the detection of quadruple images of lensed
QSO in the shape of an Einstein Cross (see
e.g. \citet{Wisotzki,Morgan} for first discoveries). However, these
optical structures are rare on the sky \citep{OM} as they require a
very close alignment of quasars with foreground massive
galaxies. Various large-area sky surveys are planned in the near
future to increase the number of such systems \citep[e.g.][]{W17,W18,
  Sch, A18} and recently the H0LiCOW (H$_o$ Lenses in COSMO-GRAILs
Wellspring) program \citep{Suyu, Bonvin} listed the five best lensed
quasars discovered to date showing an Einstein Cross structure.

During a search for high z galaxies from the Reionization Lensing
Cluster Survey (RELICS) of the Hubble Treasury Program
\citep{Salmon17}, a possible new Einstein Cross configuration around a
galaxy located at $\alpha$ =22:11:41.99, $\delta$=03:50:52.3  was discovered by \citet{Cerny} (see Figure
\ref{fig:ima}). Later on, during a search for stripped galaxies in the HST images from the RELICS project, the serendipitously re-discovery of this object by one of us (A.O.)  led to the observations presented in this article. The system, hereafter called J2211--0350, is sitting
$\sim$90 arcsecs South-West from the core of the cluster RXC
J2211.7-0349 (z=0.397) and, based on the redder color of the alleged lens galaxy,
 it was argued that the lens had to be well behind the nearby cluster. The
system is composed by an early-type red galaxy surrounded by 4 blue objects
which are arranged in the shape of a {\it "Latin Cross"} around the
central galaxy.  In Table \ref{tab:lente} we report the magnitudes and
the relative positions of the image components. Data are from the
RELICS catalogs except for the component D that we measured directly
on the HST images.

Prompted by this discovery we report here the results of optical spectroscopy of the system that allows us to confirm its lensing nature and measure the redshift of both the
lens and the source.  We adopt the concordance cosmology and assume
H$_0$ = 70 km s$^{-1}$ Mpc$^{-1}$, $\Omega_m$ = 0.3 and
$\Omega_\Lambda$ = 0.7.
  
 \section{Observations and data analysis} \label{sec:obs}

Spectroscopic observations were collected on December 1 2018 at the 10.4 m Gran Telescopio Canarias (GTC), located at the Roque de Los Muchachos
observatory, La Palma (Spain). We used the optical spectrograph OSIRIS
\citep{Cepa03} with the grism R1000B covering the spectral range 4100--7500 \AA \ , and a slit width of 1.0 arcsec. This configuration yields an effective spectral resolution of R $\sim$ 600. 

\begin{figure}
\centering
\includegraphics[width=1.0\columnwidth]{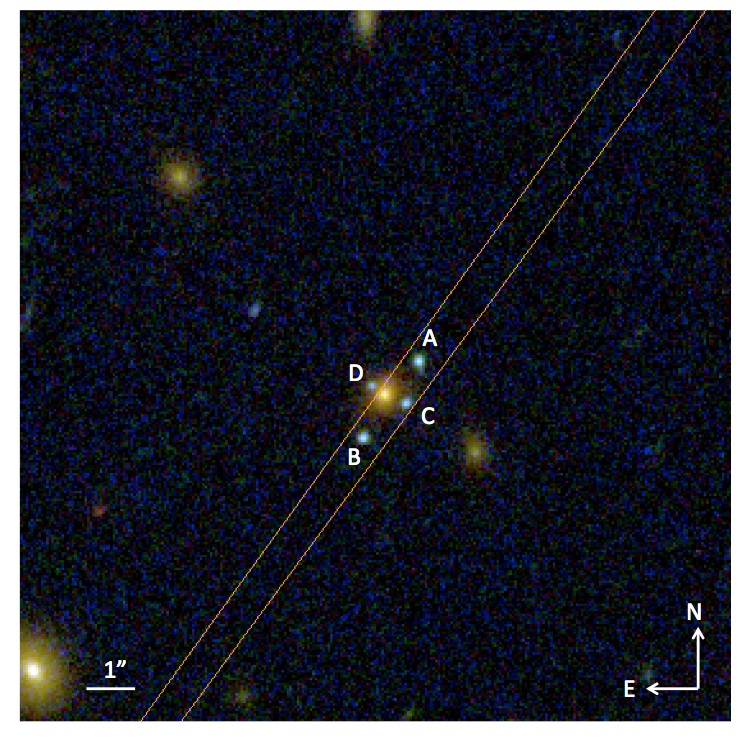}
\caption{HST color  image of the new  Einstein Cross J2211--0350 in the field of the cluster RXC J2211.7-0349.
The image is a combination of WFC3/IR (F160W) in red, ACS image (F814W) in green  and (F435W) in blue. 
The orange lines represent the position of the 1-arcsec slit used for the  spectroscopic observations.
Both the lens galaxy and three lensed targets are observed at once.}  
\label{fig:ima}  
\end{figure}

Fig. \ref{fig:ima} shows an HST image of the target with superimposed the position of the slit, located so to observe both the lensing galaxy and the three brighter images of the source at once. Only source D is outside the slit. Three independent exposures of 1800 sec were obtained under photometric conditions and
good atmospheric seeing (0.8 arcsecs).

Standard IRAF \footnote{IRAF is distributed by the National Optical
  Astronomy Observatory, which is operated by the Association of   Universities for Research in Astronomy (AURA) under cooperative   agreement with the National Science Foundation}
tools were adopted for the data reduction. Bias subtraction, flat field correction, image
alignment and combination were performed. Cosmic rays were cleaned by combining the three independent exposures and using the {\sl crreject } algorithm. The spectra were then calibrated in wavelength with
$\sim$0.2 \AA\ accuracy.  Data of a spectrophotometric standard star
observed on the same night were used to perform a relative flux calibration of the spectrum. HST photometry on filter F606W was then used to achieve absolute flux calibration of the spectrum (see Table \ref{tab:lente}).

\begin{figure*}
\centering
\includegraphics[width=2\columnwidth]{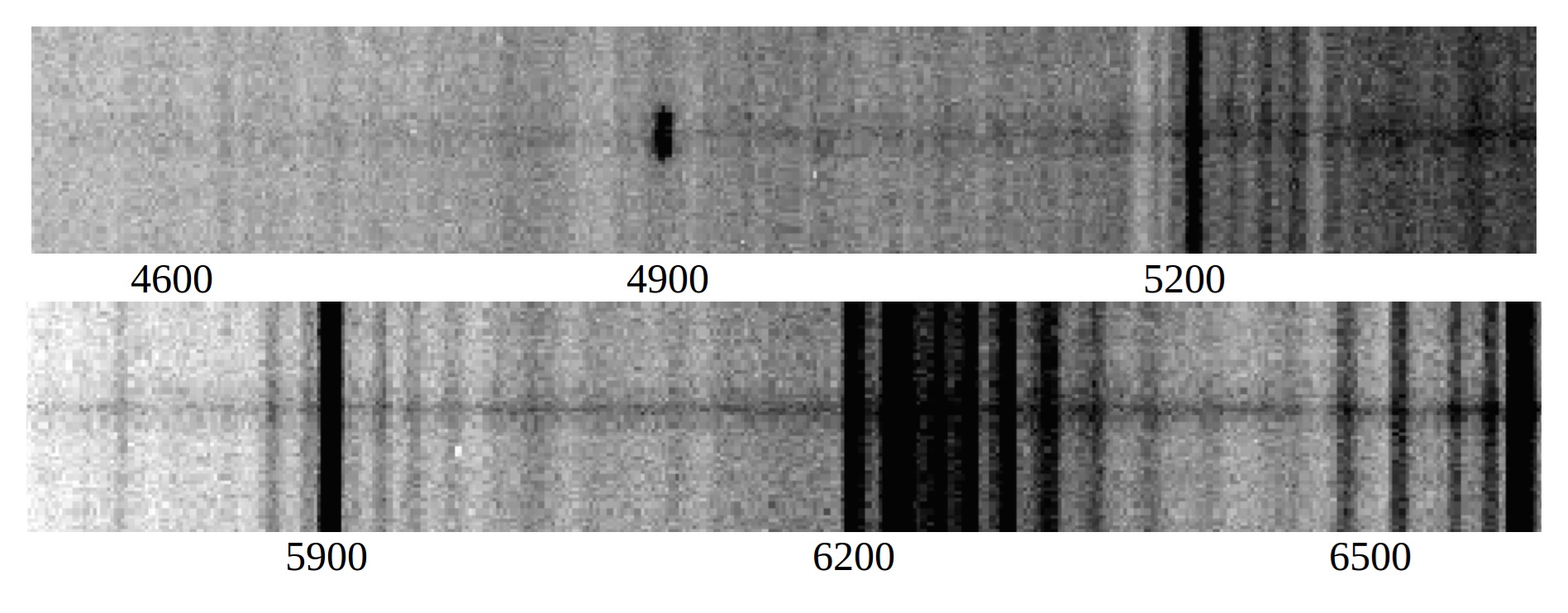}
\caption{ The 2D optical spectrum of the Einstein cross J2211--0350 (see also Fig \ref{fig:specprof}).  
Superimposed to the faint continuum is a prominent emission line at $sim$4900 \AA\, identified as Ly$\alpha$.}  
\label{fig:spec2d}
\end{figure*}


\begin{figure}
\hspace {-1.5truecm}
\includegraphics[width=1.3\columnwidth]{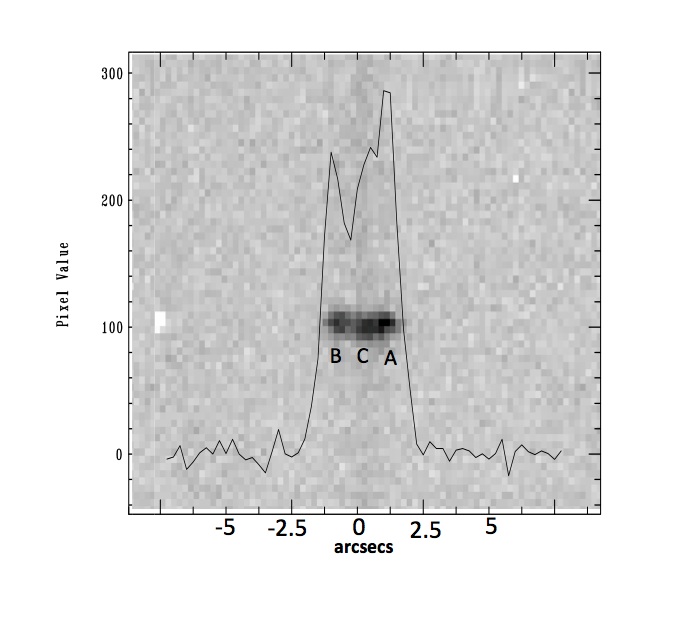}
\vspace {-1truecm}
\caption{Enlargement of the region around the Ly$\alpha$.  The   Ly$\alpha$ emission of the three lensed images (A, B and C) is   clearly resolved. Note the slight offset of component C with respect   to A and B due to the different position inside the slit (see also
  Figure \ref{fig:ima} ).  The run of the flux along the spatial   direction is also shown (solid line).\\ }
\label{fig:specprof}
\end{figure}

\begin{figure*}[h]
\centering
\includegraphics[width=2.0\columnwidth]{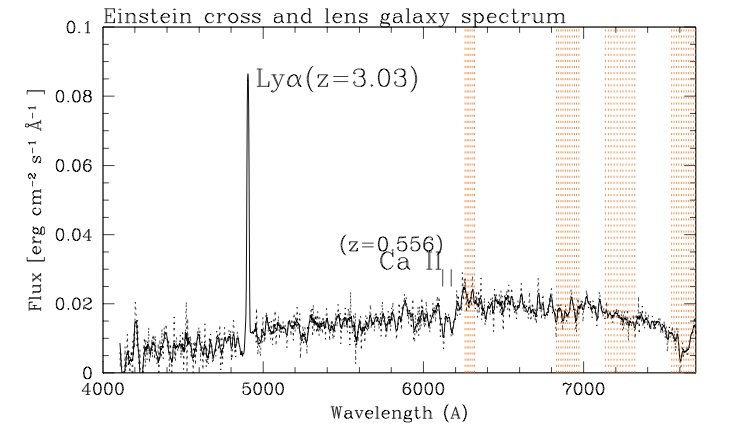}
\caption{The optical spectrum of the Einstein cross {\bf (observed frame)}. This 1D spectrum was obtained by integrating 
the flux of both the lens galaxy and three lensed images (see also Fig \ref{fig:spec2d}.
A prominent Ly$\alpha$ emission is visible at z = 3.03 due to the lensed source while stellar
absorption lines at z = 0.556 are due to the lensing galaxy.
The regions affected by telluric absorptions are marked (orange vertical lines).}  
\label{fig:spec}
\end{figure*}

The final 2d spectrum is characterized by the strong emission from
three blobs coincident with the spatial location in the slit  of the three
lensed sources (see Figure \ref{fig:spec2d} and \ref{fig:specprof}). This leaves no doubt about the nature of
this source. The emission, centered at $\lambda$=4904 \AA\, is identified as the Ly$\alpha$ at z=3.03. No other clear emission lines are visible (see Figure \ref{fig:spec}) implying that the lensed source is a Lyman Break Galaxy (LBG) rather than a QSO. The identification of this line is secure since there are no other emission lines and  its profile shows a classic, asymmetric, blue self-absorbed morphology that is typical of Ly$\alpha$ emission \citep{Jones}.

\section{Results}

\begin{figure}
\centering
\includegraphics[width=1.0\columnwidth]{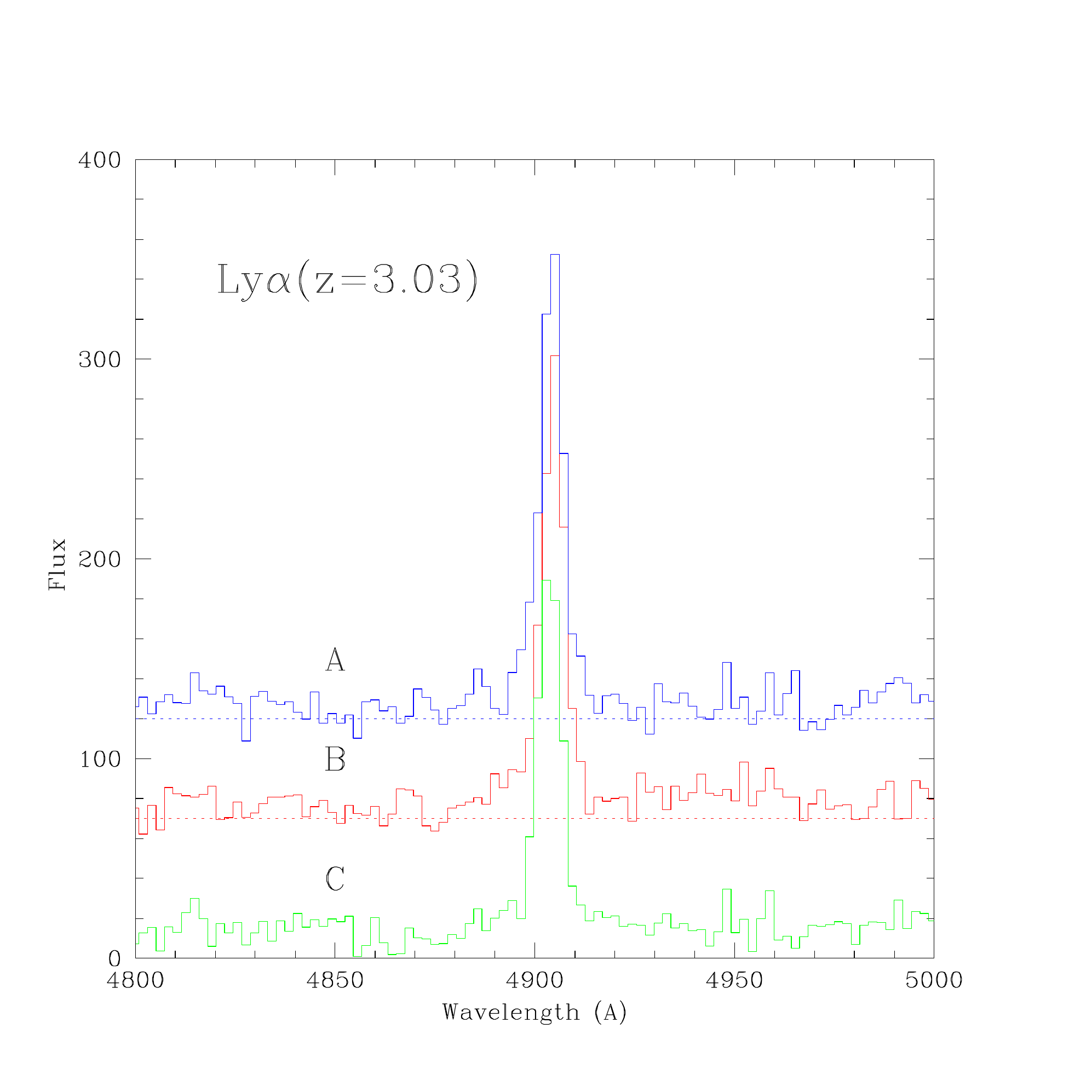}
\caption{ The Ly$\alpha$ region of optical spectrum of the three images of the Einstein cross shifted of an arbitrary quantity to viewing purposes. The peak are labeled a in Fig. \ref{fig:specprof}.  }  
\label{fig:speclya}
\end{figure}

\begin{table}[h!]

\renewcommand{\thetable}{\arabic{table}}
\centering
\hspace{-1truecm}
\caption{Position and measured magnitudes} \label{tab:lente}
\begin{tabular}{|l|r|r|r|r|r|}
\hline
  \multicolumn{1}{|c|}{Id.} &
  \multicolumn{1}{c|}{$\Delta\alpha$} &
  \multicolumn{1}{c|}{$\Delta\delta$} &
  \multicolumn{1}{c|}{F435W} &
  \multicolumn{1}{c|}{F606W} &
  \multicolumn{1}{c|}{F814W} \\
   & arcsec$^*$ & arcsec$^*$ & AB~~~~ & AB~~~~ & AB~~~~ \\
\hline
  Gal & 0 & 0 & 24.61$\pm$0.08 & 22.83$\pm$0.01 & 21.59$\pm$0.01\\
  A & 0.826 & -0.756 & 24.80$\pm$0.06 & 24.21$\pm$0.02 & 24.19$\pm$0.03\\
  B & -0.481 & 1.021 & 24.88$\pm$0.06 & 24.38$\pm$0.02 & 24.25$\pm$0.03\\
  C & 0.517 & 0.210 & 25.24$\pm$0.06 & 24.69$\pm$0.02 & 24.54$\pm$0.03\\
  D & -0.324 & -0.216 & 25.7$\pm$0.10 & 25.20$\pm$0.15 & 25.11$\pm$0.10\\
\hline
\end{tabular}\\
\small{*Positions relative to the lensing galaxy center. \\ RA(2000)= 22 11 41.97  DEC(2000) = -03 50 52.0}
\end{table}
The flux and shape of the Ly$\alpha$ emission for the three blobs are 
very similar, the observed FWHM  for this line is 10-12 \AA ~corresponding to a velocity of $\sim$ 1200
km/s at rest frame (Fig. \ref{fig:specprof}, and \ref{fig:speclya}).


The observed flux is 2.4 $\times$ 10$^{-16}$ erg cm$^{-2}$ s$^{-1}$. Including the contribution of
source D (estimated from the HST photometry), we derive a total observed line
luminosity of L(Ly$\alpha$) = 2.5$\times 10^{43}$ erg s$^{-1}$.

The red part of the spectrum is dominated by the signal from the lens galaxy. Using the pPXF \citep{CE} IDL routines we measured the stellar velocity dispersion. The CaII H and K absorption lines were fitted at z = 0.556 to estimate the galaxy velocity dispersion.  We used the library of single stellar population spectra from  \citet{Vaz} properly convolved with the instrumental resolution to fit the lines profile. The best fit corresponds to a velocity dispersion $\sigma$=208$\pm$39 km/sec. These results confirm the photometric redshift found by \citet{Cerny} an that the galaxy is not a member of the cluster RXC J2211.7-0349.


The ACS+F814W image was used to derive the properties of the lensing galaxy.
After properly masking the four images of the source, we modeled the lens galaxy with a Sersic law.
The data were deconvolved using as PSF an unsaturated and isolated star present in the field. 
We found that the lens galaxy is well fitted by a model with n=5 and effective radius r$_e$ = 0.44 arcsec
corresponding to R$_e$ $\sim$ 2.8 kpc. The absolute magnitude of the lens galaxy (z=0.556) corresponds 
to  M$_V$ $\sim$ --21.0 (taking into account k-correction). The mass of the lens galaxy estimated from the measured $\sigma$ and $R_e$ is $M_{gal}$=4.5$\times 10^{10} M_{\sun}$.
%


\subsection{Lens modeling}

We performed the lens modeling of the system by using the code by \citet{Enia}, which implements the regularized semi-linear inversion
formalism with adaptive pixel scale \citep[][and references   therein]{WD,Suyu06,NDy}.

The source plane (SP, i.e. the plane orthogonal to the line-of-sight of the observer to the deflector, containing the background source) is gridded into pixels whose values represent the
source surface brightness counts and are treated as free parameters. This approach avoids any a priori analytic assumption on the surface brightness of the background source. The size of the pixels adapts to the magnification pattern, being smaller closer to the regions of increasing magnification, to ensure a uniform signal-to-noise across the reconstructed source and to fully exploit
the increase in spatial resolution in highly magnified regions provided by gravitational lensing. For a fixed mass model of the
deflector (or lens), the SP is mapped into the image plane (IP, i.e. the plane containing the deflector and orthogonal to the line-of-sight of the observer to the deflector), then convolved with
the point spread function, and finally compared to the observed image. 
In order to avoid unphysical solutions, in with the
reconstructed source presenting severe discontinuities and pixel-to-pixel variations, a {\it regularization term} is added to the merit function. The weight of the regularization term is calculated via Bayesian analysis, according to \citet{Suyu06}.

The mass distribution of the lens is modeled as a singular isothermal
ellipsoid (SIE), described by the following parameters: the Einstein
radius ($\theta_{\rm E}$), the position of the lens centroid ($x_{\rm   L}$, $y_{\rm L}$), the minor-to-major axis ratio ($q_{\rm L}$), the orientation angle ($\theta_{\rm L}$; counter-clockwise from west). Although the object is at the edge of the nearby cluster, to model the lens, an external shear is included in the model. It is described by the shear strength ($\gamma$) and the shear angle ($\theta_{\gamma}$; counter-clockwise from west). The
search for the best-fitting parameters of the lens is done using the {\sc emcee} code (Foreman-Mackey et al. 2012), which implements the Markov chain Monte Carlo (MCMC) technique to sample the posterior probability density function (PDF) of the model parameters.

The magnification factor, $\mu$, is calculated as the ratio between the total flux density of the sources, as measured in the SP within the region of signal-to-noise ratio ${\rm SNR}\geq3$, and the flux density of the corresponding image in the IP. The uncertainty on the magnification factor is derived by computing $\mu$ 1000 times, perturbing each time the lens model parameters around their best-fitting values. \\

The modeling is carried out on the reduced ACS + F606w image, with a pixel scale of 0.06$^{\prime\prime}$. A noise map is constructed from the provided weight map and the psf is obtained by median combining 3
unsaturated stars in the vicinity of the target. The lens is subtracted from the image after fitting its light profile with GALFIT \citep{Peng}.

The best-fitting SIE model has $\theta_{\rm E}=0.76_{-0.01}^{+0.02}\,$arcseconds,
$q_{\rm L}=0.66_{-0.06}^{+0.04}$, and $\theta_{\rm L}=-35.6_{-0.8}^{+0.6}\,$degrees,
with an external shear of strength $\gamma=0.31_{-0.02}^{+0.01}$ and
angle $\theta_{\gamma}=-34.9_{-0.7}^{+0.6}\,$degrees. 
The estimated Einstein radius can be converted into a velocity
dispersion, $\sigma_{\rm SIE}$, using the relation
\begin{eqnarray}
\theta_{\rm E} = 4\pi \left(  \frac{\sigma_{\rm SIE}}{c} \right)^{2} \frac{D_{\rm LS}}{D_{\rm S}},
\end{eqnarray}
%
\begin{figure*}[t]
\centering
\includegraphics[width=1.0\textwidth]{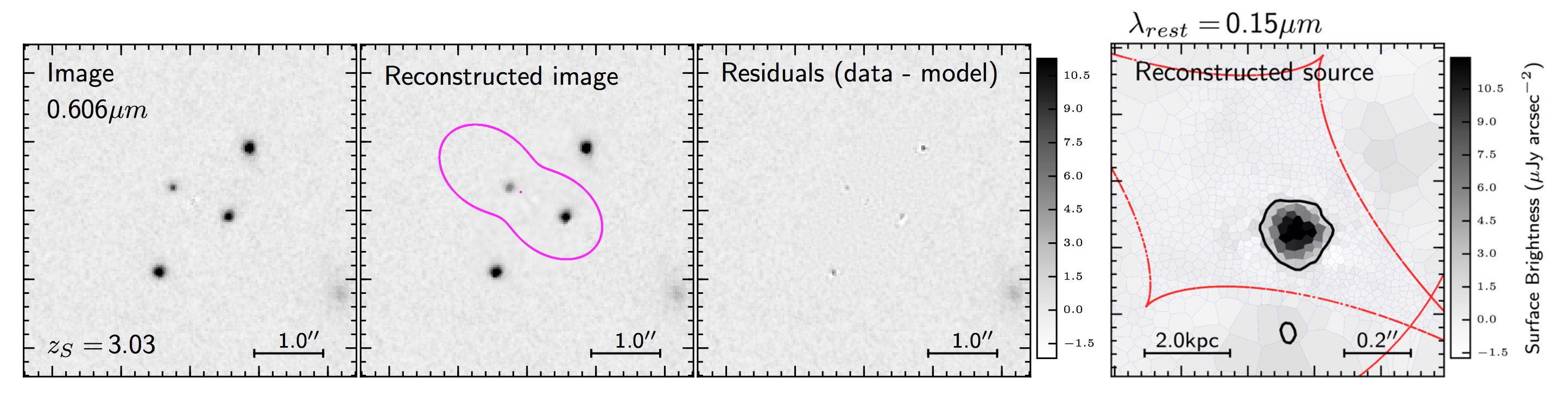}
\vspace{-0.8cm}
\caption{\small Results of the lens modeling showing, from left to right, the input image, the reconstructed image, the residuals and the reconstructed source. The purple curve in the second panel from the left is the tangential critical line, while the red curves in the rightmost panel are the caustics.}
 \label{fig:lens_modelling}
\end{figure*}

where $D_{\rm LS}$ and $D_{\rm S}$ are the angular-diameter distances from lens to source and observer
to source, respectively. We find $\sigma_{\rm SIE}=197.9_{-1.3}^{+2.6}$\,km\,s$^{-1}$, in good agreement with the value derived from the line profile.

The results of the lens modeling are shown in Fig.\,{\ref{fig:lens_modelling}}. The reconstructed source,
in the rightmost panel, is compact. This result is not surprising, as there is no evidence of extended
structure in the lensed images. The estimated magnification factor is $\mu=4.5_{-0.8}^{+1.0}$. 

\vspace{1truecm}

\section{ Summary and Conclusions}

We presented the spectroscopic confirmation of a new gravitational lens J2211 --0350 with Einstein cross configuration discovered inspecting RELICS images. The lens is an elliptical galaxy (M$_V \sim$ -21) at z = 0.556 while the lensed source  is a Ly-break galaxy at z = 3.03. Modeling of the  lens shows that the Einstein cross is well reproduced by a SIE model with Einstein ring of 0.76 arcsec and that includes the shear effect due to the foreground massive low redshift cluster of galaxies. The lensed source is magnified by a factor 4.5. 
This gravitational lens  is similar to the case reported by \citet{Bolton}  for J1011 +0143. 
This is the second case of an Einstein cross gravitational lens produced by a distant Ly-break galaxy. The intrinsic Ly$\alpha$ luminosity, taking into account  the magnification factor, is L(Ly$\alpha$) = 5 $\times 10^{42}$ erg s$^{-1}$, a factor $\sim$ 2 higher than that found by \citet{Bolton}  for J1011 +0143.  
The  Ly$\alpha$ luminosity is close to L$^*$ of the Luminosity Function of high redshift Ly$\alpha$ emitters \citep[see e.g.][]{Sobral}.

 \acknowledgments
We thank Mario Radovich, Simona Paiano and Aldo Treves for useful discussions.
Based on observations made with the GTC telescope, in the Spanish Observatorio del Roque de los Muchachos of the Instituto de Astrof'sica de Canarias, under Director$'$s Discretionary TimeÓ. This work is based on observations taken by the RELICS Treasury Program (GO 14096) with the NASA/ESA HST, which is operated by the Association of Universities for Research in Astronomy, Inc., under NASA contract NAS5-26555.

%

\vspace{5mm}
\facilities{GTC(OSIRIS), HST(ACS), HST(WFC2)}

\end{document}